# A Real-Time and Energy-Efficient Implementation of Difference-of-Gaussian with Flexible Thin-Film Transistors

Nan Wu[1], Zheyu Liu[1], Fei Qiao[1], Xiaojun Guo[2], Qi Wei[1*], Yuan Xie[3] and Huazhong Yang[1]
[1]Tsinghua National Laboratory for Information Science and Technology
[1]Dept. of Electronic Engineering, Tsinghua University, Beijing, China
[2]Dept. of Electronic Engineering, Shanghai Jiao Tong University, Shanghai, China
[3]Dept. of Electrical and Computer Engineering, University of California at Santa Barbara, USA
E-mail*: weiqi@tsinghua.edu.cn

*Abstract*—**With many advantageous features, softness and better biocompatibility, flexible electronic devices have developed rapidly and increasingly attracted attention. Many currently applications with flexible devices are sensors and drivers, while there is nearly no utilization aiming at complex computation since flexible devices have lower electron mobility, simple structure and large process variation. In this paper, we proposed an innovative method that enabled flexible devices to implement real-time and energy-efficient Difference-of-Gaussian, which illustrated feasibility and potentials for them to achieve complicated real-time computation in future generation products.**

## I. Introduction

Flexible electronic devices have attracted increasing interest, since they are soft, transparent and much more biocompatible, which leads to a broad range of potential applications, such as some future attractive and necessary products in healthcare, environmental monitoring, displays, human–machine interactivity, energy conversion, management and storage, and wireless sensor networks [1]. For example, flexible wearable electronics could be used for surveillance purposes, some researches on which have resulted in a number of prototype garments that can monitor and relay various types of data [2]. Robust and soft elastomeric devices are demonstrated in [3], which have excellent mechanical properties, for example, it could withstand pressures more than 4.0 MPa, a strain of over 110% and folding, and could be twisted more than 180°. General information about electronic skin is shown in [4], which can replace seriously damaged skin due to burst or diseases and replicate body's sense for robotic senses. Flexible electronics will also take significant market share, whose revenue can reach 30 billions USD in 2017 and over 300 billions USD in 2028 by estimation [5].

However, there are several concerns about flexible electronics. First, flexible materials always have lower electron mobility than silicon; second, flexible electronics have simple device structure and larger process variation, which indicates that it is nearly impossible to achieve real-time complex computations, such as digital image signal processing, with conventional circuit architectures. Recently most applications of flexible devices are focusing on sensors, drivers/actuators and even converters [5][6], rather than computation functions. John Rogers from UIUC proposed an idea of biostamp [7], which could be printed on the human skin and monitor body temperature, hydration and strain of skin. The sensor part is implemented with flexible materials, while the computing part is adopting integrated circuits (IC) based on silicon process, where the silicon ICs are embedded on flexible devices like dusts. In order to broaden application fields of flexible electronics, in the paper, we are trying to achieve whole computing circuits with flexible devices of thin-film transistors (TFT) to implement real-time image signal processing, which provide the potential for future whole flexible electronic systems, including sensors and complex computations.

Image filtering is a crucial preliminary step in many image processing algorithms. One of the well-known image filters is Difference-of-Gaussian (DoG) which is employed in the feature description algorithm called scaled-invariant feature transform (SIFT) [8]. Several SIFT-based algorithms are investigated in [9][10][11] as well. Since human retina can be seen as a retina kernel filter [12][13], whose general model is also a DoG, thus DoG is important in both algorithms and biological systems. In this paper, we proposed a method to implement Difference-of-Gaussian (DoG) in analog domain with flexible thin-film transistors. A Gilbert Gaussian circuit, which utilized the current-voltage characteristics in the subthreshold regime, is applied to construct Gaussian kernel, which could realized the real-time performance.

In Section II, an overview of flexible electronics and DoG is presented. We will describe circuit architecture with details in Section III and discuss experiment results in Section IV. Conclusions would be drawn in the last section.

## II. Flexible Devices and DoG

### A. Flexible Electronics

In this work, the flexible electronic devices we employed are transparent amorphous oxide (a-oxide) thin-file transistors, which have larger value of electron mobility than conventional amorphous semiconductors, such as a-Si:H, but still much less than that of silicon devices [14]. In Table I, a comparison among silicon-based MOSFET, a-Si:H TFT and a-oxide TFT is shown, from which we can figure out advantages and disadvantages of a-oxide directly. There are several advantages of a-oxide TFT devices. A-oxide TFTs have lower fabrication temperature with satisfactory operation characteristics, even if being fabricated at room temperature; moreover, they have wide processing temperature window, low operation voltage, large allowance in the choice of gate insulator, simple electrode structure and low off structure, and excellent uniformity and uniform flatness, which all indicates the ease of fabrication with a-oxide [14]. However, since a-oxide TFTs have not exhibited inversion p-channel operation, it is quite difficult to implement complex integrated designs. At the same time, with currently large process variation, large-scale ICs are nearly impossible to be taped out. The device model we applied is built based on Hspice Level 62 TFT Model [15], with refinements to adjust a-oxide properties.

As shown in Fig. 1, the settling time is 0.1 $\mu s$, which has worse performance than silicon-based devices (we chose 3.3V IO NMOS model in Hspice Level 49 as comparison) but still could do real-time signal processing with proper circuits designs.

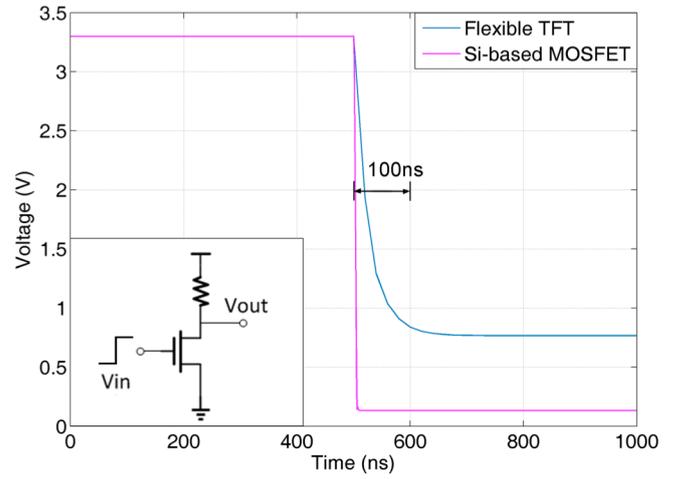

Fig. 1. Settling time of flexible devices.

TABLE I. COMPARISONS AMONG DEVICES [14][16]

| Index | Different Devices | | |
|---|---|---|---|
| | Silicon-based MOSFET | A-Si:H TFTs | A-oxide TFTs |
| Transparent | × | √ | √ |
| Soft | × | √ | √ |
| Biocompatible | × | √ | √ |
| Electron Mobility | 1400 cm$^2$/Vs | < 1 cm$^2$/Vs | 10-50 cm$^2$/Vs |
| Process Variation | Small | Medium | Large |
| Device Structure | p-type & n-type | p-type & n-type | n-type |
| Manufacture Cost/Temperature | High / High | Medium / ≤ 110℃ | Low / Room temperature |

### B. Brief Introduction of DoG Algorithm

DoG filter plays a significant role in image signal processing applications, which is also what some retinal ganglion cells carry [17]. To be more specific, DoG filter is the fundamental step of SIFT algorithm, as well as many other critical algorithms in edge enhancing and features extracting. For example, an improved DoG filter using oval recognition domain is proposed in [18], which has better face recognition and less false detection rate.

It is widely accepted that the Gaussian convolution procedure takes most computing burden of DoG [8], which is also the optimizing target with some novel circuits architectures in this work, such as physical computing [19][20][21]. Assume that the input image is I(x, y) and the Gaussian kernel function is noted as G(x, y, σ), where x and y are horizontal and vertical coordinates in pixels space, and σ is the coordinate in scale space, respectively. The output signal is the Gaussian-convoluted image, noted as M(x, y, σ), where

$$M(x, y, \sigma) = I(x, y) * G(x, y, \sigma) \quad (1)$$

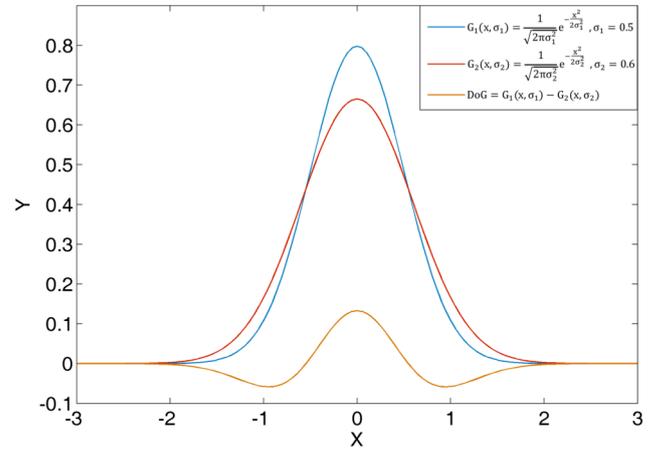

Fig. 2. Theoretical curve of Difference-of-Gaussian.

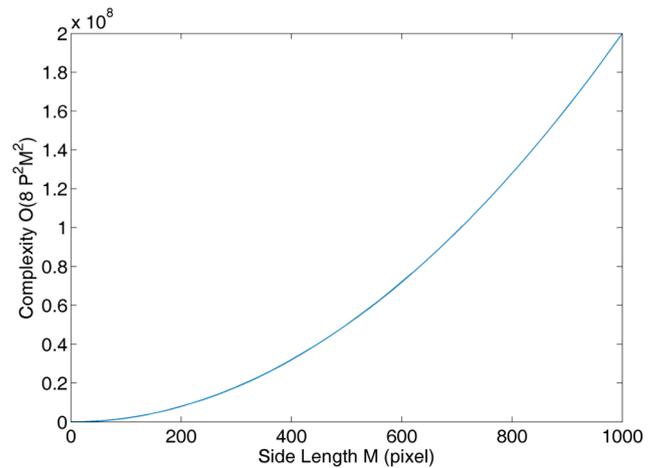

Fig. 3. Computation complexity of DoG with image size increases. Assume that M=N and P=5, and horizontal axis represents M while vertical axis is computation complexity, that is, $O(8M^2P^2)$.

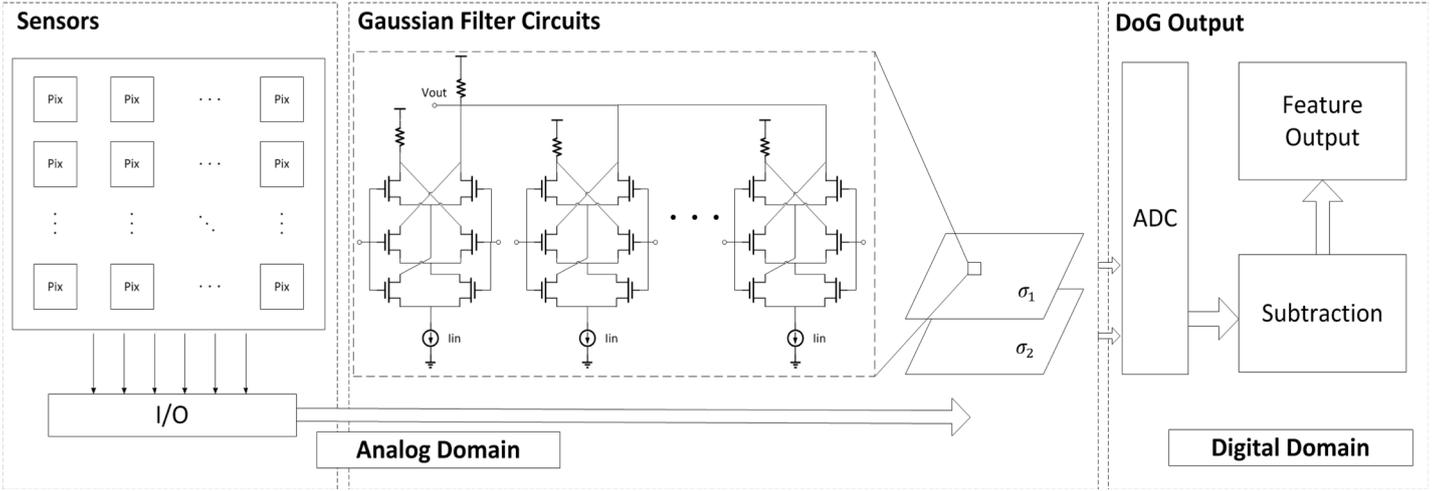

Fig. 4. Overall architechture.

and the Gaussian kernel function is

$$G(x, y, \sigma) = \frac{1}{2\pi\sigma^2} e^{-\frac{x^2+y^2}{2\sigma^2}} \quad (2)$$

As shown in Fig. 2, a DoG filter is the difference of continuous Gaussian filtered images with different scale values. For example, the difference of one image filtered by $\sigma_1$ and $\sigma_2$ respectively is

$$D(x, y, \sigma_1, \sigma_2) = M(x, y, \sigma_1) - M(x, y, \sigma_2) \quad (3)$$

For a DoG filter, if the center lobe is positive, then the parameter $\sigma_2$ should be carefully related to $\sigma_1$, and they both need relating to the kernel size of Gaussian filter [22].

Given an image with size of M×N and a Gaussian kernel with size of $(2P + 1) \times (2P + 1)$, where M, N and P are positive integers. One Gaussian filter would take $(M - 2P)(N - 2P)(2P + 1)^2$ times multiplications and $(M - 2P)(N - 2P)((2P + 1)^2 - 1)$ times additions, thus the computation complexity could be $O(8MNP^2)$, if P is much smaller than M and N. The computation complexity will increase significantly as image size increases, as that in Fig. 3, which will consume more computing time and much more hardware resource with silicon-based circuits, and would be even worse with slower flexible electronic devices.

III. PHYSICAL COMPUTING ARCHITECTURE FOR DoG

*A. Overall Achitecture*

In this section, we will present the overall system model and describe the design of circuits with flexible electronic devices, including the principle of the Gaussian circuit and the architecture of DoG circuit. As conveyed from Fig. 4, Gaussian filter circuits are integrated between photodetectors and ADC arrays, which is quite different with conventional flexible sensors. Since Gaussian filter takes most percentage computation of DoG, Gaussian convolutions are achieved with Gaussian circuits in analog domain to accelerate the process. At input source of the system, every pixel generates an analog current through illumination sensors. A Gilbert Gaussian circuit is used as a multiplier between the input current and Gaussian function, which will be detailed latter. The current-mode output in this step can be added easily by attaching those output wires together. And the eventually outputs of Gaussian filter circuits are voltages. Then the processed data passes through ADC, and subsequent procedure, such as subtraction of voltage, is done in digital domain.

Since flexible devices have large process variation currently, it is reasonable to choose relatively small size images as input signals. In order to seek more potential applications with flexible electronics, we selected MNIST datasheet and several images with simple shapes as test set, whose size is 28×28, so that we could launch more in both feature extraction and convolutional neural network in subsequent work. The proper kernel size was 3×3 counting for the chosen image sizes. The data were processed in current mode with parallel architecture and the output signals were in voltage.

*B. Gaussian Circuits*

The Gaussian filter circuit employed in the architecture is described in this section. This Gilbert Gaussian circuit, which is proposed in [23], utilizes the current-voltage characteristics of differential pairs in subthreshold regimes, multiplying two sigmoid functions. As illustrated in Fig. 5, the current-voltage (I-V) characteristic of this circuit is truly close to Gaussian curve, which can be treated as a Gaussian kernel function. The relationship between current and voltage is approximately described as (4), and for a certain device $\gamma$ is a constant.

$$I_{out} \approx \frac{I_{in}}{2} exp(-\gamma \cdot \Delta V^2) \quad (4)$$

By adjusting different $\Delta V$ with theoretical computing results could distribute various weights to different pixel.

## IV. SIMULATIONS AND EXPERIMENT RESULTS

The experiments were conducted on Hspice to analyze the performance of DoG system. In this section, an analysis about precision and processing speed is discussed.

TABLE II.  EXPERIMENT RESULTS

| Circuit Settling Time | Energy Cost | Average Deviation of Sample Points |
|---|---|---|
| 0.5 μs | 1.16nJ | 0.3341nA |

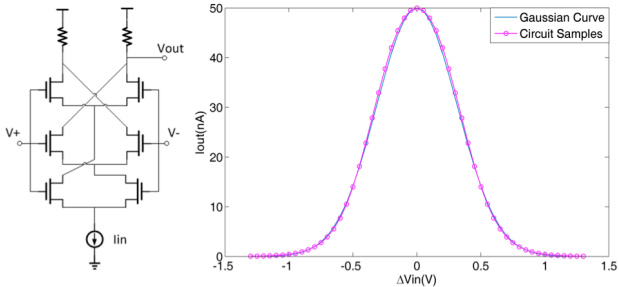

Fig. 5. Gilbert Gaussian Circuits and its current-voltage characteristic.

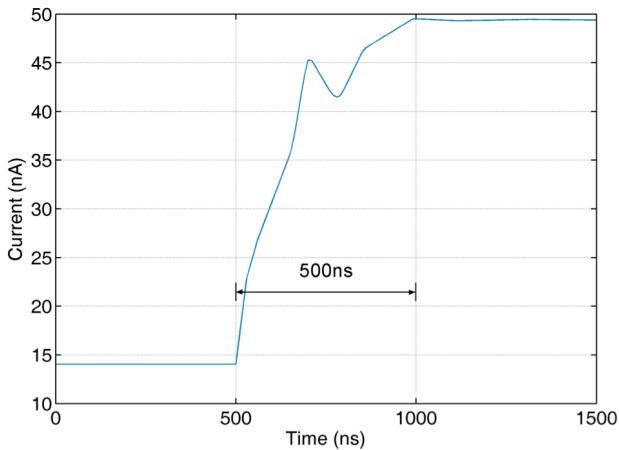

Fig. 6. Settling time of Gaussian circuits.

### A. Circuit Settling Time

The convolution speed is equal to the settling time of Gaussian circuits. According to the simulated experiment results, circuit settling time is 0.5 $\mu s$ (see Fig. 6). Assume that 200MS/s 8 bit ADCs [24] are attached to Gaussian filter output, it will take about 5ns to transform analog signals to digital signals. Since it is quite convenient to do subtraction in digital domain, the time spent could be neglected. Thus the total run time for one Gaussian convolution is about 0.5 $\mu s$, and for an image with resolution of 28×28 the total runtime is 392 $\mu s$. People can see stationary scenes as continuous figures when pictures are played in speed at least 24 frames a second, which indicates that if processing time is less than 42 ms then it could be recognized as real time process for human visual perception.

### B. Energy Analysis

In our simulation, the supply voltage is 3.3V and average current in every node is 100nA. Then the power can be calculated by $P = UIn$, where n is the total number of nodes in Gaussian filter circuits and equals 3×3 = 9 in this experiment. The result is about 2.97$\mu W$. When dealing with images with resolution of 28×28, the cost energy is approximately 2.97$\mu W$×392$\mu s$ = 1.16$nJ$. The estimated results show energy-efficiency.

### C. Precision Analysis

A single Gaussian circuit is utilized to analyze the deviation between theoretical Gaussian function and the system function of circuits with flexible devices. The average deviation of each sample point is 0.3341nA, where $I_{in}$ is set as 100nA and $\Delta V$ is scanned from -1.3v to +1.3V. (See Fig. 5)

To simplify the process and illustrate our idea more clearly, all the input data are pretreated to be binary. Since processed signals are outputted in binary mode in corresponding to input ones, subjective observation is more suitable to evaluate the quality of extracted edge features. Several typical samples and their processed results are shown in Fig. 7.

From experiment results we can find that the edges of several number are worse than that of others, that is 2, 6 and 9, which may partly due to their violent changes and thin thickness, and the limited size of kernel. In the next period, we are going to optimize circuits and architecture to pursue lesser deviation and better performance.

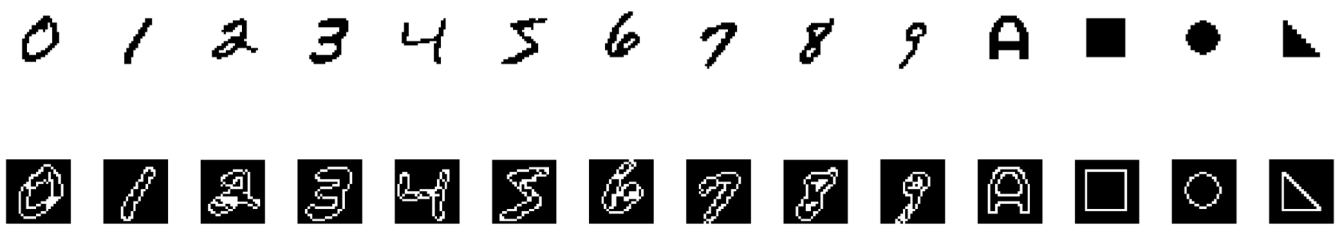

Fig. 7. Typical processed images. The upper ones are original images and the lower ones are output results.

## V. Conclusions

A novel method was proposed by this paper to make it possible for flexible devices to implement real time feature extraction. The Gaussian circuits are integrated just after sensors and accelerate signal processing in analog domain with hardware, which to some content overcomes the disadvantages of flexible devices, that lower electron mobility, simple structure and large process variation. Though there are some concerns of flexible electronics currently, it is undoubted that increasing attention would be attracted to this direction. What we have done demonstrated the feasibility of flexible devices to achieve real time and energy-efficient computation, or virtual reality if go further. Assume a condition like that, one day in the future, you buy or design ICs for specific purposes and then you could print your own ICs just with 3D printers in your home or office, rather than give them to industry and wait for a long time, since flexible devices can be fabricated in low temperature and very low cost.

At same time, there is still much work to deal with, which would seek more potential for wearable or implantable electronics with flexible devices. Firstly new materials and new devices with better performance and stability need to be introduced. Secondly some novel circuit designs with flexible electronics that enable a balance between low-power and fast speed needs to be struck. Finally, on the system level, systems should be able to process many other complex signals, such as the neural signals, which ensures implantable devices to directly communicate with bio-tissue. Many breakthroughs await people to pursue.